\relax
\documentclass[letterpaper]{article} 
\usepackage{aaai22}  
\usepackage{times}  
\usepackage{helvet}  
\usepackage{courier}  
\usepackage[hyphens]{url}  
\usepackage{graphicx} 
\usepackage{natbib}  
\usepackage{caption} 
\DeclareCaptionStyle{ruled}{labelfont=normalfont,labelsep=colon,strut=off} 
\usepackage{algorithm}
\usepackage{algorithmic}

\usepackage{booktabs}
\usepackage{comment}
\usepackage{color}
\usepackage{xcolor}
\usepackage{amsmath, array}
\usepackage{xspace}
\usepackage{placeins}
\usepackage{changepage}
\usepackage{longtable}
\usepackage{multirow}
\usepackage{siunitx}
\usepackage{supertabular}
\usepackage{enumitem}
\usepackage{textcomp} 
\usepackage{url} 
\usepackage{csquotes}

\usepackage{newfloat}
\usepackage{listings}
\floatstyle{ruled}
\newfloat{listing}{tb}{lst}{}
\floatname{listing}{Listing}
\title{Misleading Repurposing on Twitter}
\author {
    Tuğrulcan Elmas,\textsuperscript{\rm 1}
    Rebekah Overdorf, \textsuperscript{\rm 2}
    Karl Aberer \textsuperscript{\rm 1}
}
\affiliations {
    \textsuperscript{\rm 1} EPFL\\
    \textsuperscript{\rm 2} University of Lausanne\\
    tugrulcan.elmas@epfl.ch, rebekah.overdorf@unil.ch, karl.aberer@epfl.ch
}

\begin{document}

\newcommand{\Secref}[1]{Section~\ref{#1}}
\newcommand{\Figref}[1]{Figure~\ref{#1}}
\newcommand{\minisection}[1]{\vspace{0pt}\noindent\textbf{#1}}
\newcommand{\archive}{archive}
\newcommand{\Archive}{Archive}
\newcommand{\integrity}{integrity }
\newcommand{\Integrity}{integrity }
\newcommand{\snc}{screenname-changes }
\newcommand{\descr}[1]{{\bigskip\noindent\textbf{#1}}}
\newcommand{\descrplain}[1]{{\smallskip\noindent\textbf{#1}}}

\newcommand{\attr}[1]{#1}
\definecolor{cambridgeblue}{rgb}{0.64, 0.76, 0.68}
\newcommand{\bekah}[1]{{\color{blue}BO: #1}}
\newcommand{\new}[1]{{\color{cambridgeblue}#1}}
\newcommand{\positive}{\textit{positive}\xspace}
\newcommand{\negative}{\textit{negative}\xspace}
\newcommand{\unsure}{\textit{unsure}\xspace}

\maketitle
\begin{abstract}
We present the first in-depth and large-scale study of \emph{misleading repurposing}, in which a malicious user changes the identity of their social media account via, among other things, changes to the profile attributes in order to use the account for a new purpose while retaining their followers. We propose a definition for the behavior and a methodology that uses supervised learning on data mined from the Internet Archive's Twitter Stream Grab to flag repurposed accounts. We found over 100,000 accounts that may have been repurposed. We also characterize repurposed accounts and found that they are more likely to be repurposed after a period of inactivity and deleting old tweets. We also provide evidence that adversaries target accounts with high follower counts to repurpose, and some make them have high follower counts by participating in follow-back schemes. The results we present have implications for the security and integrity of social media platforms, for data science studies in how historical data is considered, and for society at large in how users can be deceived about the popularity of an opinion. 
\end{abstract}
\section{Introduction}

``As Gregor Samsa woke one morning from uneasy dreams, he found himself transformed into some kind of monstrous vermin.''~\cite{franz1915verwandlung}


Social media platforms allow users to change their profile information in order to keep up with real-world or online identity changes. For example, a user may change their real-world name and want their online identity to reflect that change, they may want to make their profile more anonymous, or they may make a career change and want to change their description field to reflect it. 

Not all attribute changes are genuine, however. Sporadically, journalists, bloggers, activists, and users report instances of accounts changing identities overnight to such an extreme degree that the former identity is lost completely, and the ``new'' account is used for a different purpose. For example, the Twitter account of an attractive woman with thousands of followers switching to an account promoting a political party, or the Twitter account of a user that reported to be based in the UK suddenly changing their name and location and taking on the identity of a patriotic American citizen. In instances such as these, accounts keep their followers but transform all of their characteristics 
at once including their \attr{name}, \attr{screen name}, \attr{description}, \attr{location}, \attr{website}, and even the style or language of the tweets. We refer to this type of drastic shift in identity and/or characteristics as \emph{repurposing}. 

Such drastic changes, in which the entire identity of the account is changed suddenly, are usually the result of malicious activity. In the most obvious case, consider a user who aims to execute some malicious activity that requires many followers, e.g., spam propagation, illicit advertisements, propaganda, political manipulation, etc. This user first creates a fake account not for its final purpose, but for the sole purpose of gaining popularity and visibility via gaining followers. That is, a new fake account posting only spam or political content will get little attention from genuine users, so a more attractive and more human-appearing account is first created to gain followers. Once this first goal is achieved and the account has risen in popularity, the account owner changes the account to achieve the intended goal. Often the final goal of the malicious user is not to use the account but to \emph{sell} the now popular account to another user who then changes it to fit their own purpose.  

In this study, we focused on this phenomenon specifically on Twitter due to data availability. However, misleading repurposing can occur on any social media platform. For example, hackers have hijacked popular YouTube channels, such as the one by Chilean urban-music artist Aisack, changed the account names and pictures to make them look like official Tesla channels, and executed bitcoin schemes~\cite{elonmuskscam}. Few platforms have explicit countermeasures to combat account repurposing, e.g., Facebook keeps a log of old names and notifies followers of popular accounts and pages of any name changes, and Reddit disallows account name changes. However, most, including YouTube and Twitter, have no such countermeasures. 

The overarching goal of this paper is to define, characterize and provide a detection methodology for misleading repurposing. Our contributions are as follows:

\begin{itemize}
\item We introduce the concept of misleading repurposing and suggest a definition. 

    \item{We present the first large-scale study of misleading repurposing.}
    
    \item We establish a hand-labeled ground-truth dataset of repurposed accounts using datasets published by Twitter.
    
    \item We provide an analysis of repurposed accounts and find that they were more likely to build towards and/or have higher follower counts. We also found that some accounts were repurposed after staying dormant for a while and deleted their old tweets.
    
    \item We propose a classifier to flag repurposed accounts in the wild. We flagged 106,000 accounts that may be repurposed in the wild by using it.
    
\end{itemize}

The structure of this paper is as follows: We first conduct a survey, motivate, and propose a definition for misleading repurposing. Then we propose a framework to find repurposed accounts. We build a dataset of repurposed accounts. We characterize the repurposed accounts. We present our classifier to detect repurposed accounts in the wild. We lastly discuss the implications of misleading repurposing.
\section{Background}
\label{sec:related}

\subsection{Survey of Related Work}
Misleading repurposing is a form of platform abuse. Adversaries repurpose accounts by misleadingly changing profile attributes. They can repurpose the accounts they already have or acquire the accounts by buying or compromising. We now give a background on such behaviors that give way to misleading repurposing.

\minisection{Platform Abuse} Platform abuse can be studied by focusing on fake~\cite{hernandez2021racketstore,rahman2019art,hernandez2018fraud}, untrue~\cite{elmas2021tactical,paraschiv2022unified}, or harmful~\cite{elmas2021dataset} posts, or misleading accounts. In this work, we focus more on a subset of misleading accounts. Past research primarily focused on accounts with automated behaviors, e.g. spammers~\cite{herzallah2018feature,danilchenko2022opinion}, fake followers~\cite{fameforsale}, impersonating bots~\cite{goga2015doppelganger}, retweet bots~\cite{elmas2022characterizing}, and astroturfing bots~\cite{elmas}. Most of the research on non-automated sock puppets relies on datasets published by Twitter, such as IRA trolls~\cite{balasubramanian2022leaders}. Our work tries to break out of this pattern. Timely detection of fake accounts through classification of repurposing behavior could lead to early detection of accounts that manipulate social media.

\minisection{Profile Attribute Changes} Profile attributes, which on Twitter consist of screen name, sometimes called handle, (\emph{@ICWSM}); name (\emph{ICWSM}); description (\emph{The 17th International AAAI Conference on...}); location (\emph{Limassol, Cyprus}); and link (\url{icwsm.org}), self-state an identity, i.e., the purported entity behind an account (e.g., a user or organization), or the purpose of the account (e.g., a hobby or fan page). Thus, users signal their social identities through such attributes~\cite{pathak2021method}. Misleading repurposing employs attribute changes to signal a change of identity or purpose. Previous work analyzed how users change their profile attributes, primarily to uncover under which circumstances these changes are made~\cite{shima2017users,neha2019change,wesslen2018bumper}. Jain et al.~(\citeyear{jain2016dynamics}) found that users change their attributes to maintain multiple accounts, change user identifiability, and for username squatting. Regarding screen names in particular, Mariconti et al.~(\citeyear{mariconti2017s}) found that adversaries hijack the screen names of popular users who recently changed their screen name in order to gain visibility, often with malicious intent. Onaolapo et al. (\citeyear{onaolapo2021socialheisting}) created fake Facebook accounts and exposed them to criminals to study how they will abuse the accounts and found that they change profile attributes to make the accounts more attractive to the victims. None of these works reported that accounts changed attributes as a signal to repurposing. 

\minisection{Accounts Changing Ownership} Repurposed accounts may have changed ownership. Accounts can change ownership either from being compromised or on mutual agreement between the previous and current account owners, often as a result of commerce. Compromised accounts are well studied in the literature, including what the compromised accounts are used for~\cite{vandam2017understanding} and user reactions to their accounts' compromisation~\cite{shay2014my,zangerle2014sorry}. Thomas et al.~(\citeyear{thomas2013trafficking}) studied how Twitter accounts were bought and sold on illicit forums. Our work builds onto these works as studying repurposing roots out compromised and/or sold accounts. 

\subsection{Survey on Terminology and Definitions}

To the best of our knowledge, there is no systematic study that defines, describes the characteristics of, or provides a method to detect misleading repurposing. However, there are a few studies, reports, or news articles related to platform abuse that reported the behavior while investigating malicious accounts. These works either did not provide a clear definition or terminology for the behavior, or they have provided a very narrow or broad definition that does not capture the behavior exactly and introduces false positives and negatives. This motivates us to survey these works, describe the drawbacks of the terminology and definitions, and formulate a clear definition.

We observe the first terms and definitions for misleading repurposing on the papers and reports that investigate suspended troll accounts that were subject to systematic platform abuse officially recognized by Twitter, which made their data available to the public. Those trolls use misleading repurposing as a strategy for platform abuse. As a result, some of the studies reported this strategy under different names which is usually the combination of ``screen-name/handle/profile/account'' and ``change/switch/hijacking/repurposing/trafficking.'' We extended our search space by searching these terms on Google, Google Scholar, and Google News as well as scientific repositories IEEE Explore and ACM Digital Library. We categorized the reports under three different names and definitions. We now survey them and indicate their drawbacks.

\minisection{Repurposing:} The term ``repurposing'' was first used by studies that analyzed IRA trolls interfering with the 2016 U.S. election. Llewellyn et al. (\citeyear{llewellyn2019whom}) used the term to describe the activity of 12 ``troll'' accounts that were suspended after working to influence the Brexit debate while posing as citizens from the U.K., though their locations were initially in Germany and their bios in German, indicating that they were repurposed.

Similarly, Zannettou et al. (\citeyear{russian_trolls}) used the term to refer to troll accounts that ``adopt different identities over time,'' i.e., they ``reset'' their profile by ``deleting their previous tweets and changing their screen name/information.'' They investigated an account that was first observed using the name ``Pen\_Air'' and later changed its name to ``Blacks4DTrump.'' Under this name, it grew its follower count from 1.2k to 9k and, after 11 months, it deleted all of its tweets and adopted the screen name ``southlonestar2,'' posing as a U.S. citizen with anti-Islamist, far-right ideology.

Both of these works provide case studies for the behavior which we study here and the latter even implicitly defines it. The definition is narrow, as it requires an account to delete their previous tweets, but we find that not all repurposings have this feature, e.g., an account that represented John Doe may be repurposed to represent Rich Roe but keep John Doe's cat videos. Thus, we develop a broader definition. 

Moreover, we also recognize that there are repurposings that are not misleading. For instance, Uren et al. (\citeyear{australian}) reported that the Chinese government used ``repurposed spam accounts'' to promote government propaganda in Hong Kong. However, these accounts underwent minimal effort to hide the fact that the accounts were repurposed (i.e., unchanged profile attributes, past tweets in a different language), therefore not explicitly misleading their followers. Thus, the term ``repurposing'' is not sufficient alone to capture such negative cases.  

 \minisection{Handle Switching:} Several reports by the Stanford Internet Observatory focusing on suspended troll accounts from the Middle East reported a repeating pattern: some accounts have at least 3 years gap between their creation date and their first (visible) tweet. They suggest that the account changed their profile handles (screen names) and wiped the old tweets, although they could not prove this behavior as they could not have access to the accounts' old handles~\cite{grossman}. Diresta et al.~\shortcite{diresta2021middle} explicitly name this behavior ``handle switching'', defined as ``growing followings, perhaps with spammy follow-back behavior, then delete old tweets and change the handle.'' 
 
This definition adds the component of growing followings to the working definition of repurposing which we explored earlier. The extra requirement makes this term too narrow for our purposes: an account does not necessarily need to grow followers to be repurposed, as is the case of the trolls discussed earlier. Additionally, the term itself refers to changing the profile handle, which also occurs in other methods of profile abuse such as handle sharing~\cite{pacheco2021uncovering} (multiple users sharing the same handle), sometimes also called URL handover~\cite{hamooni2016url}, and profile name squatting (hijacking a handle of a celebrity)~\cite{mariconti2017s}. Those cases have to be distinguished from misleading repurposing.

\minisection{Fake Account Trafficking:} Some researchers and journalists have reported on fake account trafficking, where fake accounts are purchased prior to repurposing. For instance, an account by the name of ``Oy ve Hilesi'' (English: \emph{Vote and Fraud}) was seemingly created only to attack a pro-opposition NGO called ``Oy ve Ötesi'' (English: \emph{Vote and Beyond}) during the 2015 Turkish elections. Just prior to being called ``Oy ve Hilesi,'' the account had the identity of ``a sexy girl'' and was tweeting romantic quotes as part of a scheme to artificially gain followers. Once it reached 40,000 followers, the account was sold on an underground forum. The new owner deleted the old tweets, changed the \attr{name}, the \attr{description}, and the \attr{profile picture} and, thus, shifted from a fake personal profile to an anonymous political account used to attack the opposition.
As of September 2022, the account retains 34,000 of its followers. ~\cite{sozeri}

Likewise, a pseudonymous security researcher reported an instance of several fake accounts attacking far-right French politicians in a coordinated manner. All of the accounts used the same email address and stolen photos. Later, the accounts were repurposed: they deleted all of their tweets and claimed to be ``some sort of artificial neural network company or laboratory filled with fake content.'' ~\cite{french1} 

Recently, Mazza et al. (\citeyear{mazza2022ready}) conducted a study to investigate underground markets that promote ``fake account trafficking.'' They reported that accounts that are on sale on such markets continuously change profile attributes and some are used to manipulate social media in a coordinated manner.

As trafficking fake accounts is not identical to but a means for repurposing, we proceed with the broader concept of misleading repurposing in this study. 

\section{Defining Misleading Repurposing}

\subsection{Motivation for a Broad Definition} 

Our survey yields different illicit practices that have a common component: misleadingly changing profile attributes of \emph{any number of} accounts, acquired in \emph{some way} towards \emph{some} goal. We can taxonomize misleading repurposing by three aspects to accommodate all the practices: the means, the size, and the goals. \emph{The means} is how the account is acquired prior to repurposing. It may be purchased, compromised, or the actor's own account. It may be a combination of these means, e.g., if one person created multiple fake accounts and let them age while staying idle, then another bought those accounts and used them for their illicit activities like sleeper agents. \emph{The size} refers to the number of accounts involved. A single account may be repurposed on its own or multiple accounts may be repurposed together, e.g., in a coordinated manner to boost the impact of a single narrative.
\emph{The goals} may be anything from opinion manipulation (e.g., on politics) to commercial fraud (e.g., bitcoin scams).

Although those taxonomies are useful when studying information operations on a case-by-case basis, they are redundant and often hard to detect and distinguish in the data itself. We decided to collapse these practices into what they have in common: account repurposing. Therefore, we use a broad definition of misleading repurposing to cover all cases. In other words, the accounts we investigate later may belong to different practices, but they are all subject to misleading repurposing.

\subsection{Proposed Definition}
\label{sec:define}

We present our proposed definition by introducing its components and motivating them using the negative cases from our review. We conclude with our final definition.

\minisection{Condition of Substantial Change:} The focus of this study is on accounts that undergo changes. Traditional takeovers through commercial activity or hacks are out of the scope of this paper. We are interested not in profile attribute changes generally, but in changes that make an account appear to become something else. Therefore, we require \emph{substantial changes} to the accounts, which often manifests as name and screen name changes.

\minisection{Condition of Repurposing:} Substantial changes to names and screen names by themselves do not always imply repurposing. For instance, Elon Musk changed his Twitter name to “Lorde Edge” and his location to ``Trollheim''~\cite{lordedge}. The change is more akin to a simple joke rather than a repurposing. To account for such legitimate changes, we select the term ``repurposing'' --- \emph{to adapt for use in a different purpose.} Its advantage over ``handle switching'' is that it does not encompass instances like the above example. 

\minisection{Condition of Misleading:} On its own, the term repurposing covers the relevant anomalous cases, however, it also introduces false positives. For instance, when a new president of the United States takes office, the account @POTUS is repurposed from an account of the former president to the new president and adopts the name and details of the new president. However, we do not want to consider such benign changes as interesting. The same account may have multiple purposes and/or change purposes over time. Such behavior is open and subject to public comprehension and therefore is not the focus of this study. We, thus, require that the repurposing intends to \emph{mislead} users. 

The term ``misleading'' alone is broad and subjective. Consider the following examples which we do not consider as positive cases: \emph{Anonymous} compromised the Twitter account of Burger King and changed its name to McDonald's, which signals that the account is repurposed to promote McDonald's. However, the bio of the account was ``...Just got sold to McDonald's because the whopper flopped.'', signaling that the account is temporarily repurposed as a prank~\cite{mcdonalds}. Similarly, the U.K. Conservative Party press office changed its Twitter screen name to ``FactCheckUK.'' Although their name may imply that they provide fact-checking services for the UK, their description and background image explicitly stated that they were the conservative party~\cite{factcheckuk}. In both  examples, even though the name change appears to be misleading, the account's past is evident from the current state of the account to the social media users. 

To account for such cases, we require repurposing to mislead \emph{external observers about the past of the account}. Therefore, we require misleading repurposing to have a substantial change to the account attributes so that the initial identity (i.e. the entity it represents) or purpose (i.e. what it is created for) of the account cannot be inferred from the new state of the account. Thus, the account \textit{misleads others about who they were and what they were doing}. This definition is similar to Facebook's definition of inauthentic behavior which is stated as ``misleading people or Facebook about the identity, purpose or origin of the entity that they represent'' ~\cite{inauthentic_facebook}. 

\minisection{Condition of a New, Clear Purpose:} Legitimate changes introduced to the accounts (e.g., a name change to reflect a real-world new name) can make it difficult to infer the initial identity or purpose of the account. To exclude such cases, we also require that the new state of the account have a recognizable identity or a purpose that is irrelevant to the initial identity or purpose. For personal accounts, these would mean the account has gone through an identity change and now represent a new person. For non-personal accounts such as hobby accounts centered around a subject, we adopt Facebook's definition and say ``substantial change to the account's subject.'' 

\minisection{Final Definition:} We boil down these specifications into a single definition. Thus, we define misleading repurposing as ``changing the account in a substantial way so that it represents a brand new identity or a purpose while misleading the public about the history of the account.''

\minisection{Caveat:} Misleading repurposing does not necessarily intend to cause harm, e.g., a legitimate hobby page (e.g., cat videos) can be repurposed to be used to purport another legitimate hobby page (e.g., fitness). Our goal in this paper is to uncover misleading repurposing, independent of whether the account is harmful. Uncovering misleading repurposing is a starting point in revealing harmful accounts that disrupt the public dialogue through social media manipulation.

\section{Building a Dataset of Repurposed Accounts}
\label{sec:method}

In order to study repurposing more broadly, we must first find more instances of repurposed accounts. As this is a rare event, we take a machine-learning approach to simulate the function and mass-label accounts as repurposed. We first hand-label a set of accounts as repurposed or not, then train a classifier to find more instances in the wild.

\begin{figure}[hbt]
    \centering
    \includegraphics[width=0.9\columnwidth]{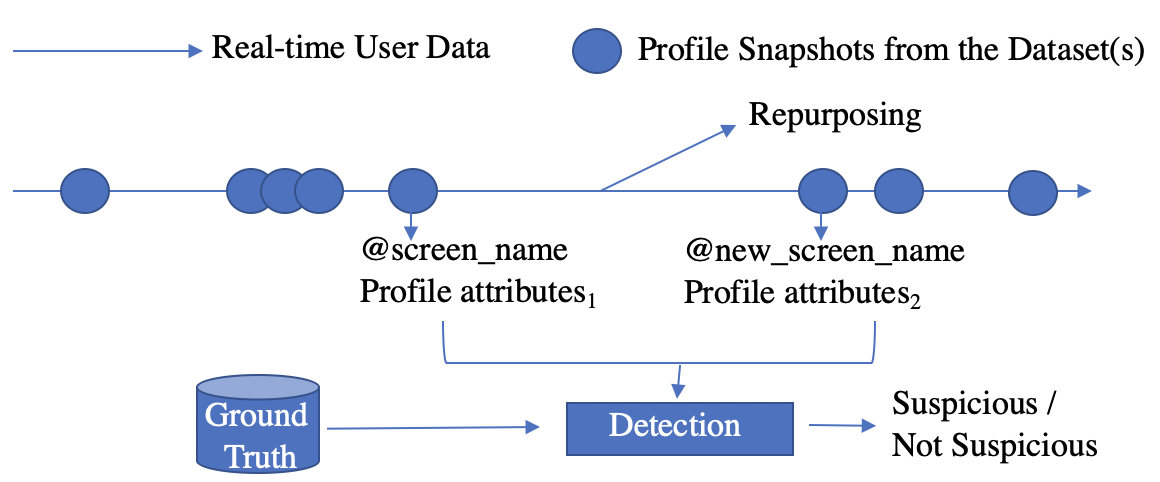}
    \caption{Summary of our methodology. We use profile snapshots of an account and detect if it is repurposed based on the ground truth we built.} 
    \label{fig:methodology}
\end{figure}

Our methodology consists of collecting a historical dataset that contains past profile snapshots to reconstruct the users' Twitter history, hand-labeling this data to establish ground truth, and building a classifier to detect suspect repurposed accounts in the wild. The process is summarized in Figure \ref{fig:methodology}.

\subsection{Base Dataset (Archive)}

\label{sec:data} 
In order to detect if an account has been repurposed, we must, at the very least, have a snapshot of the account before the repurposing and a snapshot after to witness the change. To this end, we use a dataset of public Twitter data that is archived by the Internet Archive's Twitter Stream~\cite{team2020archive}. This dataset contains a 1\% sample of all tweets, including retweets, and includes a profile snapshot of the user who posted the tweet and, if applicable, the retweeted user. As the sample includes retweets, popular users with many retweets will appear more often than accounts with fewer of these interactions. Thus, the dataset is biased towards active and popular accounts, which is an advantage in this study, as such accounts have the greatest potential to have an impact. We name this dataset \emph{archive}.

Even though it contains only 1\% of tweets, this dataset is massive. At the time of analysis (October 2020), the dataset dated from September 2011 to June 2020 and contained 446 million \attr{user id}s. We create an abbreviated version of this dataset by only considering accounts that changed their screen name since we found this signal in every example of repurposing we studied. In general, this is a rare action on Twitter since a screen name provides a URL for the profiles so that they can be searched and shared (i.e. \url{twitter.com/justinbieber}) so the users may prefer to keep it the same. Additionally, Wesslen et al.~(\citeyear{wesslen2018bumper}) found that this attribute is stable for most users. We found that only 13.3\% (59 million) of users in the archive dataset changed their screen name, confirming this finding.

\subsection{Ground Truth Datasets}

To establish ground truth, we hand-label a set of accounts. Our preliminary analysis shows that randomly sampling Twitter users who changed their screen names to find positives is not an efficient strategy as repurposing is extremely rare. Additionally, negative cases that are randomly sampled are very trivial to classify: they are often slight changes to screen names with little or no changes to names and description fields. Therefore, we follow a multi-step approach: we first hand-labeled a collection of accounts from a set that we know contains repurposed accounts and build a simple classifier on this data with sufficient precision and recall. We then deploy this classifier \emph{in the wild} (i.e., on normal Twitter accounts) and detect positives. We then manually annotate these positives and create a second ground-truth dataset.

\minisection{Civic-Integrity Ground Truth Set (Integrity)} We use the datasets published by Twitter that involve state-sponsored accounts that undermine election integrity for the first step. Some were already reported to be repurposed by previous work. By October 2020, there were 35 datasets focused on 16 countries~\cite{twitter_infoops}. The datasets do not include past profile attributes. Thus, for each user id in these datasets, we extracted the historical data from the archive dataset. Of the 83,481 unique \attr{user ids}, 38,426 were found in the \archive. We found 17,220 screen name changes involving 8,370 accounts. We name this dataset \emph{integrity}.

\minisection{In The Wild Popular Users Ground Truth Set (Popular)} We have a ground truth set of accounts with many positive cases. However, this set is biased. We observed that malicious accounts often change their name and description field drastically for the purpose of misleading repurposing. This makes the probability of misleading repurposing given the drastic change in profile attributes close to one, making two events seem the same. Additionally, the base rate of the positive cases in the integrity dataset is very high compared to a random sample. As a result, in our preliminary experiments, we observe that our classifier reporting good scores on the integrity dataset performed poorly in the wild and yield many false positives. Thus, we used the accounts detected by our initial classifier as a ground truth set for the second step. We then took an active learning approach and deployed new and more complex classifiers to improve our initial one. We only include the accounts with at least 5,000 followers (see Ethics section). We name this new ground truth set the \textit{popular} dataset.

\section{Annotation}

We use human annotation to build a ground truth for repurposed accounts. We refrain from crowdsourcing this task because 1) the \integrity dataset is only fully available to researchers given access by Twitter; 2) the archive dataset, although public, contains sensitive information, e.g., former names of real users; and 3) expert annotation may be more reliable than crowdsourcing for complex tasks.


We treat each instance of a screen name change as a separate data point. For each change from screen name $s_i$ to $s_j$, we select the last available snapshot of the profile with $s_i$ and the first snapshot with $s_j$. The same user may have multiple screen name changes and, thus, may be represented multiple times. We presented the annotators with the following semantically interpretable attributes: \attr{name}, \attr{screen name}, \attr{description}, \attr{location}, \attr{home page url}, \attr{profile settings language}, most common \attr{tweet source} and \attr{tweet language}. We asked the annotators the following question and allowed for the responses: \emph{Yes(+)}, \emph{No(-)}, and \emph{Unsure}.

\begin{displayquote}
\emph{Did the account change in a way that makes it seem that the account is now owned by a different person/organization, or has the account rebranded itself substantially?} 
\end{displayquote}

Two authors/domain experts, $A_1$ and $A_2$, independently coded all cases. We report the annotator agreement using Cohen's kappa. We additionally hired a third non-expert student annotator, $A_3$, who was not involved in the research process at any stage. We report the annotator agreement for the three annotators using Fleiss $\kappa$. Especially for the \emph{popular} dataset, we found that the expert annotators had a higher agreement amongst themselves than with $A_3$. 

Due to the subjective nature of the problem, we observed many cases in which we needed to code and determine a common answer. The relatively low agreement of a non-expert further emphasized this need. Since this problem is unexplored, there is no coding scheme available. Thus, $A_1$ and $A_2$ developed a coding scheme and made decisions for differently coded cases when needed. First, each annotator independently annotated the data using only the initial annotation question. Then they compared the annotations and computed the annotator agreement. They then discussed and coded the cases in which they disagreed or were both \unsure. Finally, a decision was made for each case. Below, we present the cases and the decisions. 

\subsection{Annotated Data}
\minisection{Integrity Dataset} We first selected English and French profiles for validation by multiple annotators. $A_1$ and $A_2$ independently annotated 200 cases. $A_1$ labeled accounts in this sample and passed 100 \positive and 100 \negative cases to $A_2$ for annotation. The inter-annotator agreement between the authors was $\kappa$ = 0.8 (substantial agreement). For all three annotators, Fleiss $\kappa$ = 0.79.  
To expand this labeled dataset, $A_1$ labeled an additional 1,476 profiles in English, French, and Turkish (Turkish accounts were made available after the initial annotation was complete). This resulted in 512 \positive, 910 \negative, and 254 \unsure.

\minisection{Popular Dataset} For the popular dataset, we used a stratified sampling approach and sampled 400 accounts from the list of users with the most followers before the repurposing and 600 accounts from a random sample of users who had more than 5,000 followers. Half of those accounts tweeted in English while the other half tweeted in Turkish.

For the popular dataset, the annotation was done simultaneously: $A_1$ and $A_2$ independently annotated the samples. This resulted in  $\kappa=0.66$ including \unsure cases (i.e., one decided \negative while the other decided \unsure was considered disagreement) and  $\kappa=0.81$ when cases decided as \unsure were discarded from the data. For all three annotators, Fleiss $\kappa=0.46$ (moderate agreement) including \unsure cases and $\kappa=0.63$ (substantial agreement) excluding \unsure cases. We observe that the disagreements were mostly due to overestimating the prevalence of repurposings as all accounts in this dataset substantially change their name and descriptions.

$A_1$ and $A_2$ then discussed the cases in which they did not agree and either came to a consensus or assigned a label of "Disagree" in the case of disagreement or \unsure if both annotators were \unsure of the case. This annotation resulted in 562 \positive cases, 127 \negative cases, 278 \unsure cases, and 33 disagreed cases. 

To expand this labeled dataset, $A_1$ additionally annotated 1,500 accounts with the same sampling strategy. This annotation was done more conservatively and the goal was to increase \negative examples, since in-the-wild \negative cases with dramatic name changes are rare. Only the cases where the annotator was highly confident were annotated as \positive; \negative and \unsure cases were  not checked for a second time. This yield an additional 421 \positive, 248 \negative, and 831 \unsure. 


Our coded cases and decisions for each are presented in Table~\ref{tab:annotation}. We explain each in detail.

\begin{table}[ht]
\centering
\caption{Summary of Cases in Our Annotation Framework}
\label{tab:annotation}
\begin{tabular}{p{50pt}|p{130pt}|p{30pt}}

Case & Example & Verdict \\\hline\hline
New Identity/Subject & John Doe, a lawyer, becomes Mohammad Lee, a doctor & Positive \\\hline
Commercial Activity & Account named ``sdfsf" or ``FOR SALE" becomes John Doe & Positive \\\hline
Same \hspace{10pt} Person & Jane Doe marries and becomes Jane Brown & Negative \\\hline
Purpose Overloading & Jenna The Traveler becomes politics enthusiast Jenna Abrams & Negative \\\hline
Slight Ch. In Subject & Philosophy Quotes becomes Inspiring Quotes & Negative \\\hline

No Purpose / Unclear & White Horse becomes Black Rose but still shares quotes & Unsure \\\hline
Org.\hspace{25pt} Rebranding & Windows Phone becomes Lumia & Unsure \\\hline
Non-Substantial Ch. & Patriotic Somalian changes name, but keeps description & Unsure \\\hline
Person$\leftrightarrow$Org. Unclear & John Doe becomes ”Lonely Boy’s Pen” & Unsure\\\hline
Change\hspace{5pt}pseudoynms& Excalibur17 becomes Rebellion47 but still plays DOTA & Unsure
\end{tabular}
\end{table}


\subsection{Positive Cases}

\minisection{New Identity or Subject:} The account purports a completely new person, hobby page, or organization when compared to its old version. The account has a new name, a new website, and a new location. It is easy to infer the purpose of the old snapshot and the new snapshot from the description, and they are dramatically different. 

\minisection{Commercial Activity:} The old purpose is unclear because the account is either blank and named by a random string (which indicates fake account trafficking), or it self-reports that it is for sale. We assume commercial activity has taken place and the new owner repurposed the account.

\subsection{Negative Cases}

\minisection{Same Person:} A user changed their profile attributes but it is evident that they are the same person. This is often apparent because keywords are shared between snapshots or the writing style of the account does not change. We observe that many teenage pop-artist fans change their attributes frequently to express their admiration in different ways. We annotate these cases as \negative if we can confidently infer that the account purports the same person and purpose. Otherwise, we annotated them as \unsure.

\minisection{Purpose Overloading:} The account appears to be the same person, but it is changed in a way that it is repurposed to share a new type of content, e.g. politics. We identified 43  troll accounts originating from Russia that appear to be personal profiles that had initially politically neutral description fields but then later adopted description fields that exhibit their political stance (e.g., pro-Israeli, patriot, conservative, \#Blacklivesmatter) alongside a corresponding change in demographic attributes (e.g., Christian, Black). This is a stealthy and potentially malicious strategy, but it is not a case of misleading repurposing. 

\minisection{Slight Change in Subject Matter:} Although the profile attributes have changed, the subject matter did not change substantially and it appears to be the same. 

\subsection{Unsure Cases}

\minisection{No Purpose or Purpose Unclear:} It is difficult to understand the purpose of the account. It was either because the annotators did not have the cultural context although could understand the language (e.g. Nigeria) or the lack of profile attributes that state the purpose of the accounts.

\minisection{Organization Rebranding:} Some organizations were sold or rebranded. Examples include musical.ly which became TikTok, Windows Phone which became Lumia, and Facebook which became Meta. We consider those as \unsure since it is not clear if they are repurposed nor if it is misleading. Other than those obvious examples, it is difficult to distinguish between a rebranding and a new company without thorough research. Thus, if the purpose of the previous and the next snapshot of the organization is the same or similar even though they appear to be different organizations, we annotate such organizations as \unsure.  

\minisection{Non-Substantial Changes} Adversaries change profile attributes but the changes are minimal, e.g. the name of the account change but the description field stays the same. In those cases, either it is not possible to judge if the accounts purport the same people or if the change is misleading. 

\minisection{Person $\leftrightarrow$ Organization Unclear:} An account that has the same purpose, but it is repurposed to be a page or an organization when it was a personal profile. It is not clear if this should be considered misleading repurposing because it could be the same person being professional or adopting a pseudonym (fictitious name) for their hobby. We annotate such cases as \unsure. Exceptionally, if a profile appears to be a user with a hobby turning their personal page into a hobby page, we consider it \negative. E.g., John Doe stating that he shares photographs adopts the name DoePhotography.

\minisection{Change of Pseudonyms:} It is not clear if a person/page is repurposed to be a new person/page even though the name and description changed dramatically. These people/pages did not change their domain. We observe the former among esports gamers as they sometimes switch pseudonyms and teams, but they play and stream the same game. We observe the latter among meme pages as their names and description fields are also memes but there is no other indication of the specific purpose of the account.

\section{Characterization}

We describe some of the characteristics of the accounts that have undergone misleading repurposing. 
Specifically, we discover that misleading repurposed accounts often 1) have more followers than other accounts that change their screen names, 2) utilize follow-back schemes to grow their follower counts, 3) delete tweets related to their former purpose, and 4) have a period of dormancy before the repurposing.

\begin{figure}[hbt]
    \centering
    \includegraphics[width=\columnwidth]{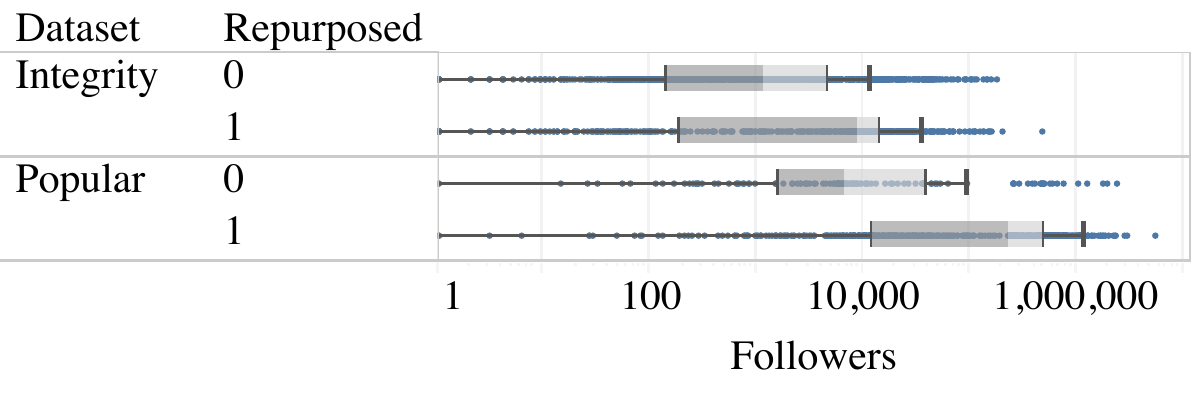}
    \caption{A box plot showing the number of followers for repurposed vs not repurposed accounts. High follower counts are more likely to indicate repurposing. } 
    \label{fig:followers}
\end{figure}

\minisection{Follower Count} Accounts in our dataset with a high number of followers are more likely to be misleadingly repurposed when they change their screen name than other accounts that undergo a screen name change. Screen name changes are an anomaly for influential accounts since they lose any incoming links (i.e., if @jack changes his screen name then twitter.com/jack does not redirect to Jack's new screen name). Similarly, high-follower accounts may be more likely to be the target of account swapping due to being compromised or due to commercial activity. This may be an artifact of the data collection: users with a high number of followers are more active so it is more likely that we capture their screen name changes and, thus, their repurposing. Similarly, this does not imply that misleading repurposing is more prevalent among accounts with high follower counts. Figure \ref{fig:followers} illustrates this difference.

In the integrity dataset, the mean followers count before the screen name change was 14,007 for repurposed accounts and 6,579 for non-repurposed accounts. The difference is 7,427 and statistically significant according to Welch's t-test. In the popular dataset, the mean followers count before the screen name change is 327,431 for repurposed accounts and 139,237 for non-repurposed accounts. The difference is 188,194 and statistically significant according to Welch's t-test ($p < 0.0001$ in both cases). 

\minisection{Follow-Back} Many repurposed accounts appear to actively grow their accounts by joining follow-back schemes. They indicate that they follow back once another account follows them by using dedicated hashtags. Out of 1,595 repurposed accounts, 81 accounts used \#FF ("Follow Friday", which is the most used hashtag in the dataset), 50 accounts used \#Follow, 44 accounts used \#IFollowBack, and 36 accounts used \#TeamFollowBack. Meanwhile, out of 1,385 non-repurposed accounts, only 9 accounts used \#FF, 7 used \#TeamFollowBack, and 5 used \#Follow. Our caveat is that these numbers are based on 1\% of the tweets and there may be many more users using those hashtags. 

\begin{figure}[hbt]
    \centering
    \includegraphics[width=\columnwidth]{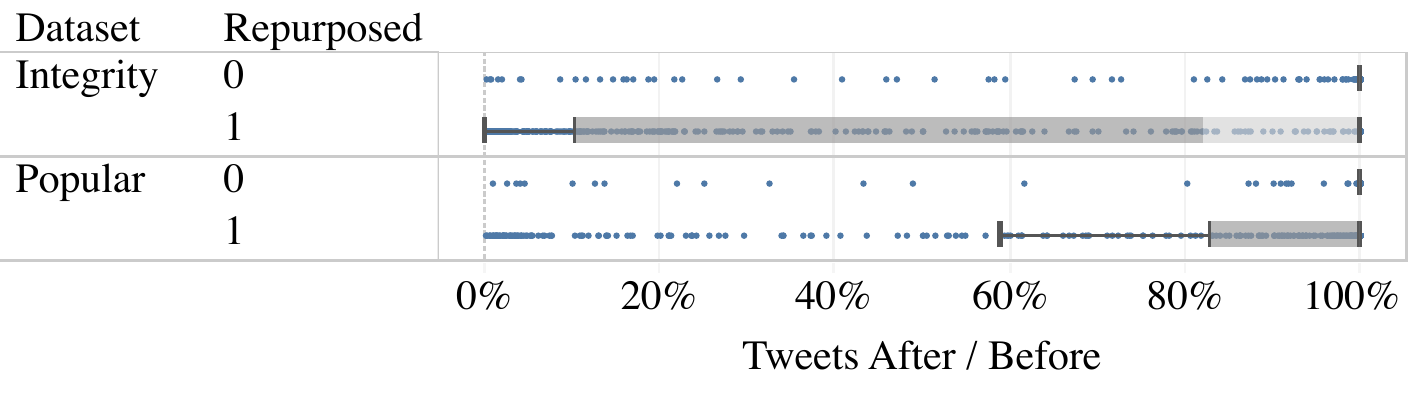}
    \caption{Box plot of the ratios of the number of tweets before and after an account changed its screen name for repurposed vs. not repurposed accounts. Accounts that created more tweets than deleted are not included in the plot for visualization purposes. Repurposed accounts are more likely to delete their tweets when they change screen names.} 
    \label{fig:deletions}
\end{figure}

\minisection{Deletions} Repurposed accounts often delete tweets that are irrelevant to the new purpose of the account. We observe this behavior by comparing the number of tweets before and after the account changed its screen name. ~\Figref{fig:deletions} shows that the repurposed accounts are more likely to lose up to 96\% of their tweets. Precisely, 519 of the 1,595 repurposings (32\%) resulted in removing at least one tweet versus 75 of the 1,385 non-repurposings (5\%). The difference is statistically significant according to the chi-squared test with $p < 0.0001$.

\minisection{Dormancy} We observe that repurposed accounts in the integrity datasets are more likely to be dormant for a long period prior to repurposing. This may be because the owners of the accounts no longer use them and eventually sell them. Alternatively, the accounts get compromised but since the original owner does not use them, they do not claim them and let them be repurposed by another malicious user. We did not observe this behavior among popular accounts.

\begin{figure}[hbt]
    \centering
    \includegraphics[width=\columnwidth]{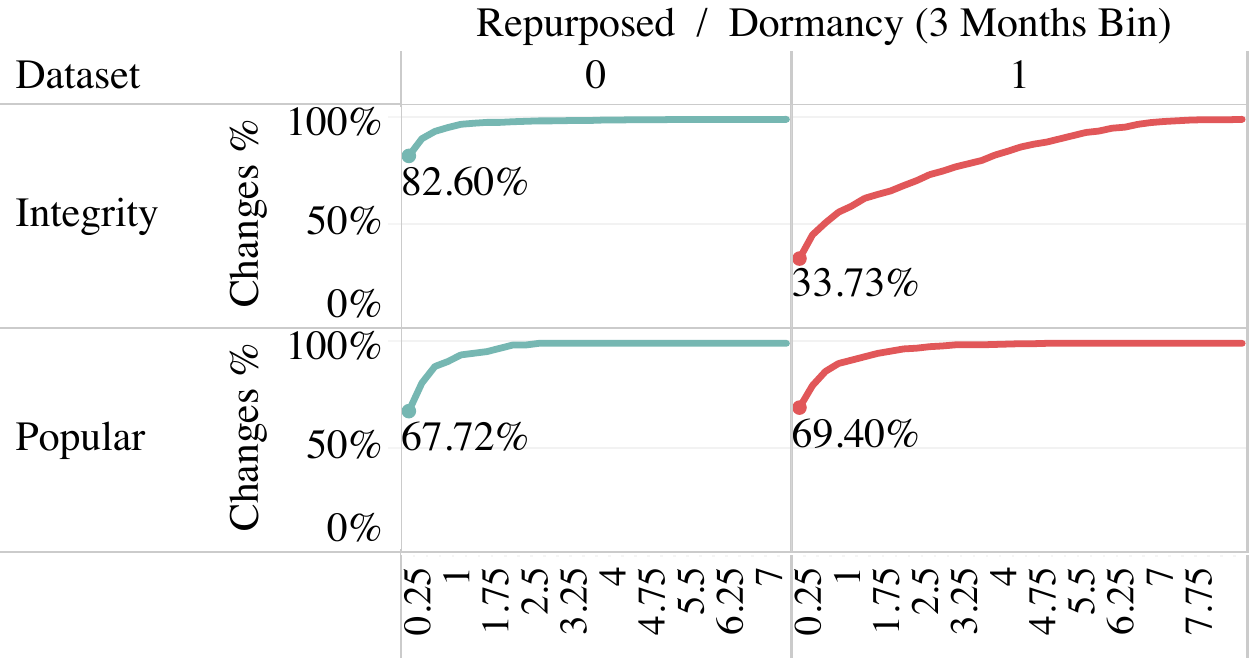}
    \caption{Cumulative distribution (CDF) of the percentage of accounts staying dormant. Bins are 3 months/quarter years. Misleading repurposed accounts in the integrity dataset are more likely to repurpose after staying dormant for a while, unlike popular accounts.} 
    \label{fig:dormancy}
\end{figure}

\section{Detection}

\begin{table*}
\caption{Results on the integrity dataset (-I) and the popular dataset (-P). Best performances in bold. We use F1 as the primary evaluation metric for the integrity dataset and AUC as the primary evaluation metric for the popular dataset due to distinct base rates. We report the other scores for completeness.}
\label{tab:results}
\begin{tabular}{l|l|l||l|l|l||l|l|l|l|l}
\toprule
            Model &     F1-I &  AUC-P &   Prec-I &    Rec-I &    AUC-I & Prec-P &  Rec-P &   F1-P &  TPR-P &  FPR-P \\
\midrule
    EDT (Baseline) &  92.3\% &      - &  94.4\% &  90.3\% &  92.6\% &  81.5\% &      - &      - &      - &      - \\
        \textbf{EDT-DSIM} &  92.8\% &  \textbf{88.4\%} &  95.5\% &  90.3\% &  97.3\% &  92.1\% &  92.7\% &  92.4\% &  92.7\% &  37.4\% \\
         EDT-STY &  93.1\% &  73.2\% &  91.7\% &  94.6\% &  97.5\% &  88.9\% &  87.3\% &  88.1\% &  87.3\% &  51.4\% \\
             EDT (Retrained) &  94.0\% &  78.5\% &  95.6\% &  92.5\% &  97.8\% &  88.7\% &  92.3\% &  90.5\% &  92.3\% &  55.1\% \\
          EDT-MD &  94.0\% &  79.4\% &  95.6\% &  92.5\% &  98.4\% &  90.9\% &  89.3\% &  90.1\% &  89.3\% &  42.1\% \\
     \textbf{EDT-DSIM-MD} &  94.0\% &  \textbf{87.6\%} &  95.6\% &  92.5\% &  98.6\% &  91.5\% &  92.1\% &  91.8\% &  92.1\% &  40.2\% \\
 EDT-DSIM-MD-STY &  94.5\% &  84.8\% &  96.6\% &  92.5\% &  98.1\% &  92.3\% &  90.5\% &  91.4\% &  90.5\% &  35.5\% \\
      EDT-MD-STY &  94.6\% &  76.4\% &  95.6\% &  93.5\% &  98.2\% &  91.0\% &  82.1\% &  86.3\% &  82.1\% &  38.3\% \\
    \textbf{EDT-DSIM-STY} &  \textbf{95.1\%} &  83.1\% &  96.7\% &  93.5\% &  97.5\% &  91.2\% &  92.3\% &  91.7\% &  92.3\% &  42.1\% \\
\bottomrule

\end{tabular}
\end{table*}

We next provide a classifier to detect misleading repurposing in the wild. The goal of this detection method is not to develop the framework that should be used by Twitter or other social media companies to detect repurposed accounts, as they have access to a richer set of signals and data. Instead, we provide a framework for researchers who do not have such privileged access to flag accounts that are potentially repurposed by only using publicly available data. Due to the subjective nature of the problem, we advise that the detection should always be accompanied by expert verification. We tackle the following classification problem:

\minisection{Problem Statement} 
Let $\mathcal{A}$ be an account with $n$ snapshots $\{A_{t_0}, ... A_{t_n}\}$. Let $A_{t_i}=\{scn_i,P_i,\tau_i\}$ where $scn_i$ is the screen name of $A_{t_i}$, $P_i$ is the profile information of $A_{t_i}$, and $\tau_i$ are all of the tweets currently available on $A_{t_i}$. For each pair $A_{t_i}$ and $A_{t_{i+1}}$ where $scn_i \not = scn_{i+1}$, determine if the changes to $scn_i$, $P_i$, and/or $\tau_i$ indicate a misleading repurposing. It follows that a negative result is an account that has not been misleadingly repurposed, either because it has not been repurposed or it was not repurposed in a way that misleads users about its past states. 

We experiment with four classification strategies based on different features to measure changes to $scn_i$, $P_i$, and/or $\tau_i$: change of name and description, name/name similarity, profile metadata, and style change. We use a combination for the final classifier.

\minisection{Change of Name \& Description (EDT)} We observe that changing the name and description thoroughly at the same time with the screen name is a behavior that is indicative of misleading repurposing. Thus, we create features to capture this signal based on edit distances. We compute the Levenshtein distance between the string fields and use it as a feature. The formula is as follows:

\begin{equation}
    NLD_\text{attr} = \frac{\text{lev}(\text{attr}_{U_{\text{prev}}}, \text{attr}_{U_{\text{next}}})}
    {max(len(\text{attr}_{U_{\text{prev}}}), len(\text{attr}_{U_{\text{prev}}}))} 
\end{equation}

\noindent where attr is the string attribute, \textit{len} is its length, $U_{\text{prev}}$ is the previous snapshot, $U_\text{next}$ is the next snapshot.

We adopt an online learning approach and train a simple classifier using this feature to sample accounts from the popular dataset. Precisely, we train a decision tree classifier of depth two, which classifies screen name-changing instances with \textit{$NLD_{\text{name}} > 0.721$} and  \textit{$NLD_{\text{description}} > 0.742$}. 

We chose this classifier because it initially achieved sufficient precision and recall. Thus, we use this classifier as \textbf{the baseline} and build other classifiers to improve on it. 

\minisection{Name/Description Similarity (DSIM)} 
After training the initial baseline classifier, deploying it in the wild, and finding more positives and negatives, we made an observation: non-malicious users who thoroughly change their name and description field leave some artifact that is relevant to the past of the account in order not to mislead their audience, e.g. old screen names, email addresses. We compute features to exploit this behavior. We compute the longest common sequence between the two snapshots' names, screen names, and description fields combined to account for the longest common substring. We computed the raw number and Jaccard coefficient of common tokens between two texts to identify common entities. Finally, we compute the similarity between those two texts using sentence-transformers. We use the model "bert-base-multilingual-uncased"~\cite{DBLP:journals/corr/abs-1810-04805} since our data consists of different languages.

\minisection{Profile Metadata (MD)} We employ additional textual features such as the home page of the profile, the self-stated location, and the profile image. We check if these attributes changed and also compute the edit distance and their $NLD$ (except for the profile image). We also introduce non-textual (numeric) profile attributes: friends count, followers count, statuses count, and favorites count. For each profile attribute in each snapshot, $S_i$ and $S_j$, we use the raw numbers, $a_i$ and $a_j$; the difference, $a_i - a_j$; and the ratio of the difference and the maximum to capture the magnitude of the change, $(a_i - a_j) / max(a_i, a_j)$. 
We also introduce dormancy which is the time passed between two snapshots.

\minisection{Style Change Detection (STY)} If misleading repurposing occurs, the style of the tweets may change because the ownership of the account may have changed. We create features based on this assumption using state-of-the-art style change detection techniques ~\cite{iyer2020style,zhang2021style} and the model ``bert-base-multilingual-uncased" \cite{DBLP:journals/corr/abs-1810-04805}. We concatenate the tweets before and after the change of the screen name and treat them as separate \textit{paragraphs}. Iyer et al.~\shortcite{iyer2020style} predict the style change between two consecutive \textit{paragraphs} by averaging their sentence vectors. We produce the sentence vectors by exactly following their method: we split each paragraph into sentences and generate embeddings for each sentence. This results in a tensor with 12 x l x 768 dimensions, where 12 is the number of layers, 768 is the hidden size and l is the length of the sentence (maximum 512 tokens). We first sum the embeddings of the last 4 layers, producing tensors of size l x 768. We then sum this tensor over the first axis to produce a vector of size 768. We generate these vectors for every sentence in each document representing the tweets posted before and after the screen name change and sum them. We then take the average of the two vectors.  
\subsection{Classification}

We train each classifier on the data annotated by only $A_1$. It consists of 512 positives and 910 negatives from integrity data and 421 positives and 248 negatives from popular data. We have 933 positives and 1,158 negatives in total. 

We experiment with several supervised machine learning algorithms: SVMs, Logistic Regression, Naive Bayes, Decision Trees, Random Forest, and Neural Nets using sklearn~\cite{pedregosa2011scikit}. We experimented with different parameters using grid search. While choosing the best model, we use the integrity dataset as the validation dataset and report the model for each classification strategy that performs the best on this dataset. We use the F1-score to evaluate the performance as the dataset is balanced. We found that Random Forest yielded the best scores consistently, so we only present the results of the Random Forest classification. As we noted before, misleading repurposing is very prevalent in the integrity dataset and even very simple classifiers perform well. Therefore, we test our classifier using the popular dataset. The goal is then to decrease the False Positive Rate among Popular accounts (FPR-P) while still sustaining a high True Positive Rate (TPR-P). Thus, we use Area Under the ROC curve (AUC-P) to evaluate our classifiers' performance. This metric is more reliable than Precision, Recall, and F1-Score when the dataset is imbalanced and positives are more prevalent as it takes FPR into account~\cite{bekkar2013evaluation}.

\subsection{Results}

All results are presented in Table \ref{tab:results}. We observe that the EDT classifier (based only on the change of name and description field) performs well on the integrity dataset but not on the popular dataset. This indicates that screen name-changing behavior generally entails misleading repurposing in the former dataset while not necessarily in the latter. The extra features based on the name and description field greatly improve this simple classifier because for most cases in which name and description changes are not due to misleading repurposing, those fields are either semantically similar or have some traces referring to old snapshots of those fields. Profile metadata features that represent the characteristics of the accounts contribute to the performance of the integrity dataset but only slightly boosts the popular dataset.

The style change classifier performs poorly on its own, suffering from a very low recall (57\% on integrity accounts and 35.5\% on popular accounts). Most of the true positives in the integrity dataset come from the accounts originating from China. As such, it improves the combinations of other classifiers when used on the integrity dataset but is not as useful when used on the popular dataset. The popular sample only contains 1\% of tweets from each user, so it is quite possible that with more data on each account this classifier would perform better. Style change may be more effective in the presence of more data. We leave a comprehensive style change analysis on social media for future work. 

The performance of EDT-DSIM is only slightly higher on English accounts compared to Turkish accounts ($AUC = 88.1 $ vs $AUC = 87.8$). It performs better on the accounts with the most followers compared to random accounts ($AUC = 89.8 $ vs $AUC = 88.0$). This may be because repurposing is more evident in accounts with the most followers as they are more likely to self-state their purpose.

\minisection{Miss-classifications} We manually examined the false negatives and false positives introduced by the BASE-DSIM classifier. False negatives occur in two cases. First, the account leaves the description field empty, leaving an insufficient amount of information for the classifier. As the popular dataset is collected using the baseline classifier, we only observe this among the integrity accounts. The annotators could annotate those accounts as repurposed due to changes in their names signifying a new person/organization. One limitation of our approach is that it relies on non-empty description fields. However, because accounts with blank description fields are rare, as we discuss below, this limitation is not critical. Second, the classifier captures common slogans and phrases as similar such as ``Follow us" and ``Updates about x". Such similarities may indicate that the owner of the account is the same or the new owner keeps the style but does not entail the absence of misleading repurposing. A special case is that the purpose of the account changes and it is misleading, but the owner appears to be the same and has the same specialization, so it continues to use buzzwords like Deep Learning. This is generally the case when the personal account becomes an organization account.

False positives occur mainly in two cases. Case 1: the classifier fails to capture the similarity apparent to an annotator, e.g, a religious page that adopts different names and quotes different religious texts with the purpose of sharing religious quotes. Since there are no repeating texts and the semantic similarity is fairly low compared to other examples, the classifier classified this example as positive. Case 2: there is enough information for an annotator but not for the classifier. This occurs if, e.g., the description field is empty in one of the snapshots but an annotator can judge that the purpose of the account has not changed based on the name alone.

\minisection{Estimation} We deployed our classifier on the 1.57 million popular accounts that were active in the first half of 2020. We estimate 180,689 misleading repurposings by 106,548 accounts. By May 2022, 22,063 (20.7\%) were suspended and 7,800 (7.3\%) were deleted. The suspension rate may be low because repurposing is not explicitly against Twitter rules.

\section{Conclusions and Implications}
\label{sec:discussion}
In this work, we provide a definition and detection methodology for misleading repurposing. Our analysis shows that certain signals such as changes in profile attributes, tweet deletions, and building a follower base via follow-back schemes may signal misleading. We also show that detecting the repurposings reliably is possible but it is subject to data availability, which poses a challenge, especially for the platforms where retrospective data is not available. We conclude our study by discussing its implications. 

\minisection{On Security and Integrity} Many social media platforms asses user credibility via user features (account creation date, number of followers, engagements, etc.). Repurposing makes it possible for a new and perhaps untrustworthy user to repurpose an older, more trustworthy account, therefore passing through the low-quality content filter. 

Furthermore, repurposing aids in coordinated manipulation by allowing for more efficient use of fake accounts. Coordinated inauthentic activity is unanimously explicitly prohibited by social media platforms. However, repurposing accounts with different histories poses a challenge for detection, as the owner of the accounts may opt to purchase accounts from different sources and repurpose them together decreasing the likelihood that the set of accounts was used in a coordinated manner in the past. As such, repurposing allows malicious actors to use the same, limited, set of accounts for multiple operations before detection.

Finally, misleading repurposing can harm user trust in the platform. First, from the perspective of a follower of a repurposed account, they find themselves following an account with content that they did not intend to follow. Second, social media users use follower count as a proxy for value to the community (e.g. popularity or interest). By repurposing an account that already accumulated a high number of followers, they manipulate this proxy.

\minisection{On Data Science} Misleading repurposing poses a challenge for studies using social media data. Such studies often assume data is static and do not consider distinguishing between the time the data is targeted, collected, analyzed and the results are reported. The data of repurposed accounts are dynamic: the accounts may reset their profiles during a course of a study. For example, an actor employs multiple accounts to push a harmful narrative. Researchers observe this activity in the wild and collect the users promoting that narrative. Then the actor repurposes all of the accounts, deletes the activity, and resets them to make them look like harmless profiles. After, the researcher will not be able to retrieve the past activity of those accounts to investigate and may not be able to infer if they were automated and used in coordination with other accounts. The accounts may still cause harm as they have already propagated the harmful narratives, but the researchers will not be able to investigate such impact, as was the case in~\cite{grossman}.

\minisection{Public Impact} Misleading repurposing is one of the sophisticated techniques adversaries employ in spreading their harmful narratives on crucial topics such as elections and public health. For instance, IRA trolls used repurposing in combination with other techniques to interfere with the 2016 U.S. elections and Brexit~\cite{llewellyn2019whom}. Both votes were decided by a very close margin, thus, flipping the opinion of a small number of voters may have changed the outcome, although it is yet to be definitively shown that such activities impacted the outcome. 

Misleading repurposing has also been used to spread dangerous rumors. A repurposed account from Saudi Arabia with 90k followers, renamed ``@QtrGov,'' spread a rumor of a coup attempt in Qatar with a fake video of gunshots~\cite{grossman}. It received 2,300 interactions and other accounts promoted the rumor. It was later debunked and reported in the news media~\cite{aljazeera}. We investigated this account using our framework and found that before deleting its old tweets and adopting the name @QtrGov, the account was aggressively increasing its followers by promising follow-backs (stated in the description) and only sharing Islamic quotes, perhaps a strategy to attract a religious audience. Misleading repurposing helped in exploiting the large follower base the account built, bringing the rumor to more users and the public's attention.

Misleading repurposing has also had financial impacts. Accounts repurposed on YouTube to imitate Tesla and Elon Musk have launched bitcoin schemes, with a single account scamming users out of almost a quarter million USD in just one week. \cite{elonmuskscam}

Finally, misleading repurposing fuels underground economies. Mazza et al. \shortcite{mazza2022ready} tracked a sample of fake accounts sold in underground markets and found that some accounts were later repurposed and used in coordinated campaigns to promote politicians in Argentina, sports products while posing as football fans, and cryptocurrency scams while impersonating popular users. Misleading repurposed enabled these fake accounts to be used in coordination while reducing the risk of suspension en masse, making it more appealing to purchase existing accounts.

\section{Ethical Implications}
\label{sec:ethics}
\minisection{Data Collection and Management} This study only uses public data provided by Twitter and the Internet Archive, both of which have been analyzed extensively by previous work. To comply with the Twitter Terms of Service and protect the privacy of Twitter users, we do not share the data of repurposed accounts from the popular dataset. However, we will share the code and the ids of the repurposed accounts from the integrity dataset, since these accounts have already been made public by Twitter and, as such, there is no risk of further harms in their release.

\minisection{Threats to User Anonymity and Privacy}
We additionally mitigate any privacy loss to normal Twitter users by limiting our study to only two types of accounts: 1) accounts in the civic integrity dataset which have been designated by Twitter as harmful to public dialogue and released by Twitter, and 2) popular accounts which can influence the public. For an account to be considered ``popular'', we follow Twitter's lead in choosing a threshold of 5,000 followers, the threshold Twitter uses in the civic integrity dataset to determine if a user's profile be made public. This group of accounts does include legitimate users who do not intend to mislead others or participate in malicious activity, and in the course of our study, we uncovered their former account names/old profiles via parsing publicly available data. This may include accidental deanonymization of a currently pseudonymized account if the user self-stated their identity in an old version of their profile and posted enough tweets from the old version of their account to appear in the 1\% sample. We mitigated this risk to the best of our availability by not releasing the data publicly, performing the annotation ourselves to not expose the data to crowd workers, and not reading their tweets.

\minisection{Further Potential Impacts of Work} We must also consider the impact of publishing such a study and making this type of platform manipulation known to the general public and academic community. First, we hope that this work raises awareness among Twitter users that accounts that they follow may be repurposed for malicious purposes so that they can notice such accounts when they see them, and possibly even report them as malicious. We also hope that pointing out and studying this phenomenon urges academics and Twitter alike to put more resources into mitigation methods that do not have negative impacts on normal users, especially those from already marginalized groups. 

Awareness goes both ways, though, and this paper could also lead to malicious users learning about repurposing. This could lead to some who did not know that repurposing was possible to maliciously repurpose more accounts. However, we know from the widespread use of malicious repurposing that this phenomenon is already known by many who wish to use it maliciously. By bringing this problem to light, we hope to mitigate this risk by promoting user and platform awareness, thus discouraging its use. 

Although the goal of this paper is to uncover malicious repurposing, parts of our methodology could be repurposed to deanonymize users who want to remain anonymous, as long as at one point in the past their account had an identifiable attribute. Users should be made aware that if they wish to remain anonymous, a new account should be created from scratch rather than repurposing a non-anonymous account.

Finally, this work further illustrates that deletion privacy is important for users, but that it also can prevent malicious activity from being discovered. While users need to be able to delete and hide their prior activities and accounts, this study underlines how such mechanisms can be misused to mislead and deceive users.

\fontsize{9.0pt}{10.0pt} \selectfont
\bibliography{bib}
\bibliographystyle{aaai}

\end{document}